\documentclass{ws-procs9x6}

\def\citebk#1{[\hspace{0.9mm}\raisebox{-1.85mm}[0mm][0mm]
  {\Large\cite{#1}}\hspace{-0.1mm}]}

\begin{document}

\title{%
Nonperturbative solution of supersymmetric gauge theories%
$^*$%                                                            %preprint only
}

\author{J.~R.~Hiller}

\address{%
Department of Physics \\
University of Minnesota-Duluth \\
Duluth MN 55812 USA \\
E-mail: jhiller@d.umn.edu}

\maketitle

\abstracts{%
Recent work on the numerical solution of supersymmetric
gauge theories is described.  The method used is SDLCQ
(supersymmetric discrete light-cone quantization).  
An application to $N=1$ supersymmetric Yang--Mills theory in
2+1 dimensions at large $N_c$ is summarized.  The addition of 
a Chern--Simons term is also discussed.}%
\footnotetext{$^*$Preprint UMN-D-02-4,                           %preprint only
to appear in the proceedings of
the fifth workshop on Continuous Advances in QCD (Arkadyfest),
Minneapolis, Minnesota, May 17-23, 2002.}

\section{Introduction}

Although much has been learned about supersymmetric gauge theories
by analytic methods, numerical methods can yield much more of
the nonperturbative structure.  In particular, the method known
as supersymmetric discrete light-cone quantization 
(SDLCQ),\cite{Sakai,SDLCQreview}
an extension of ordinary DLCQ,\cite{PauliBrodsky,DLCQreview} has
been quite successful in the analysis of (1+1)-dimensional supersymmetric
theories.  This work has recently been extended to 2+1 
dimensions\cite{SYM2+1a,SYM2+1b,T2_2+1,BPS2+1} with consideration 
of $N=1$ supersymmetric Yang--Mills (SYM) theory, including
a Chern--Simons (CS) term.\cite{CSreview}  
The mass spectrum, Fock-state wave functions,
and a stress-energy correlator are all computed.  The CS term provides
an effective mass that reduces the tendency of SYM to produce stringy
states with many constituents.  This work was done at large-$N_c$, but
the method is also applicable to finite $N_c$.

As the name SDLCQ implies, light-cone coordinates\cite{Dirac} are used.
They are defined by spacetime coordinates
\begin{equation}  \label{eq:coordinates}
x^\pm=(t\pm z)/\sqrt{2}\,,\;\;{\bf x}_\perp=(x,y)  
\end{equation}
and momentum components
\begin{equation}  \label{eq:momentum}
p^\pm=(E\pm p_z)/\sqrt{2}\,,\;\;{\bf p}_\perp=(p_x,p_y)\,. 
\end{equation}
The dot product of two such four-vectors then becomes
\begin{equation} 
p\cdot x=p^+x^-+p^-x^+-{\bf p}_\perp\cdot{\bf x}_\perp\,.
\end{equation}
The $x^+$ direction is treated as the direction of time evolution,
which makes the conjugate variable $p^-$ the light-cone energy.
The light-cone three-momentum is $\underline{p}\equiv(p^+,{\bf p}_\perp)$.
In a frame where the net transverse momentum ${\bf P}_\perp$ 
is zero, the mass eigenvalue problem becomes
\begin{equation}  \label{eq:EigenProb}
2P^+P^-|P\rangle=M^2|P\rangle\,,
\end{equation}
where $|P\rangle$ is also an eigenstate of three-momentum 
$\underline{P}$.
The SDLCQ method provides a means to solve this eigenvalue
problem with supersymmetry preserved exactly at any level
of the approximation.

One of the advantages of light-cone coordinates is
that there exists a well-defined Fock-state expansion
for each mass eigenstate.  There are no disconnected 
vacuum pieces, because the longitudinal momentum of
each constituent, virtual or real, must be positive.
SDLCQ uses a Fock-state expansion for $|P\rangle$ to
obtain a matrix eigenvalue problem for the Fock-state
wave functions at discrete values of the momentum.
The matrix is then diagonalized by appropriate means.
For large matrices the Lanczos diagonalization technique\cite{Lanczos}
has been used, as discussed in Ref.~\citebk{SYM2+1b}.

The discretization is accomplished by restricting the
fields to periodic boundary conditions in a light-cone 
box\cite{PauliBrodsky,DLCQreview}
defined by $-L_\|<x^-<L_\|$ and $0<x,y<L_\perp$.
This leads to a discrete momentum grid
\begin{equation} p^+\rightarrow\frac{\pi}{L_\|}n\,,\;\;
{\bf p}_\perp\rightarrow(\frac{2\pi}{L_\perp}n_x,\frac{2\pi}{L_\perp}n_y)\,.
\end{equation}
The product $P^+P^-$ is independent of $L_\|$, and
the limit $L_\|\rightarrow\infty$ is exchanged for a limit
in terms of an integer $K$, called the harmonic resolution,\cite{PauliBrodsky}
defined by
\begin{equation}  
K\equiv\frac{L_\|}{\pi}P^+\,.  
\end{equation}
Longitudinal momentum fractions $x=p^+/P^+$ then reduce to $n/K$.
The number of particles in a Fock state is limited to K, because
negative longitudinal momentum is not allowed and the individual
integers $n$ must sum to $K$.\footnote{Zero modes are ignored.}
The eigenvalue equation (\ref{eq:EigenProb}) becomes a coupled set of 
integral equations for the Fock-state wave functions in which the
integrals are approximated by discrete sums over the momentum grid
\begin{equation} \int dp^+ \int d^2p_\perp f(p^+,{\bf p}_\perp)\simeq
   \frac{\pi}{L_\|}\left(\frac{2\pi}{L_\perp}\right)^2
   \sum_{n,n_x,n_y}  f(nP^+/K,2{\bf n}_\perp\pi/L_\perp)\,. 
\end{equation}
The harmonic resolution provides a natural cutoff for $n$.  The
transverse sums must be truncated explicitly, which is done by 
limiting $n_x$ and $n_y$ to range from $-T$ to $T$.  The integer
$T$ can be viewed as a transverse cutoff or, at fixed dimensionful
cutoff $\Lambda_\perp\equiv 2\pi T/L_\perp$, as the transverse
resolution.

The distinction between DLCQ and SDLCQ lies in the choice of
operator for discretization.  In ordinary DLCQ one discretizes
the Hamiltonian, $P^-$; in SDLCQ, one discretizes the
supercharge $Q^-$ and constructs $P^-$ from the superalgebra
relation
\begin{equation}
\{Q^-,Q^-\}=2\sqrt{2}P^-\,,
\end{equation}
which guarantees that the discrete eigenvalue problem 
preserves supersymmetry.\cite{Sakai,SDLCQreview}  The
$P^-$ of ordinary DLCQ differs from the supersymmetric $P^-$
by terms which disappear in the large-$K$ limit but which
break the supersymmetry at finite $K$.

The remainder of this paper is organized as follows.  In
Sec.~\ref{sec:SYM}, (2+1)-dimensional SYM theory is reviewed
and numerical results discussed for the spectrum and 
for a correlator of the stress-energy tensor.  Section~\ref{sec:SYM-CS}
extends the study of the spectrum to include the CS term, in both a 
dimensionally reduced theory and the full (2+1)-dimensional case.
A brief summary is given in Sec.~\ref{sec:summary}.

\section{SYM$_{2+1}$ theory} \label{sec:SYM}

\subsection{Formulation}

The action for $N=1$ SYM theory in 2+1 dimensions is
\begin{equation}
S=\int dx^+ dx^- dx_\perp \mbox{tr}(-\frac{1}{4}F^{\mu\nu}F_{\mu\nu}+
{\rm i}{\bar\Psi}\gamma^\mu D_\mu\Psi)\,,
\end{equation}
with
\begin{equation}
F_{\mu\nu}=\partial_{\mu}A_{\nu}-\partial_{\nu}A_{\mu}
              +ig[A_{\mu},A_{\nu}]\,, \quad \quad 
D_{\mu}=\partial_{\mu}+ig[A_{\mu},\quad]\,.
\end{equation} 
The fermion field is separated into chiral projections
\begin{equation}
\psi=\frac{1+\gamma^5}{2^{1/4}}\Psi\,,\qquad
\chi=\frac{1-\gamma^5}{2^{1/4}}\Psi\,,
\end{equation}
only one of which is dynamical.  In light-cone gauge, $A^+=0$, 
with the transverse component of the gauge field $A_\perp$ 
written as $\phi$, the action becomes
\begin{eqnarray}\label{eq:action}
S&=&\int dx^+ dx^- dx_\perp
\mbox{tr}\left[\frac{1}{2}(\partial_-A^-)^2+
D_+\phi\partial_-\phi+ {\rm i}\psi D_+\psi+ \right.  \nonumber \\
& &
\left. \hspace{15mm} 
  +{\rm i}\chi\partial_-\chi+\frac{{\rm i}}{\sqrt{2}}\psi D_\perp\phi
    +\frac{{\rm i}}{\sqrt{2}}\phi D_\perp\psi \right]\,.
\end{eqnarray}
The non-dynamical fields $A^-$ and $\chi$ satisfy constraint equations
\begin{equation}
A^-=
\frac{g}{\partial_-^2}\left(i[\phi,\partial_-\phi]+2\psi\psi\right)\,, \quad
\chi=-\frac{1}{\sqrt{2}\partial_-}D_\perp\psi\,, 
\end{equation}
by which they can be eliminated from the action.
The dynamical fields are expanded in terms of creation operators
\begin{eqnarray}
\phi_{ij}(0,x^-,x_\perp) &=& \frac{1}{\sqrt{2\pi L_\perp}}\sum_{n^{\perp} = -\infty}^{\infty}
\int_0^\infty
       \frac{dk^+}{\sqrt{2k^+}}\left[
       a_{ij}(k^+,n^{\perp})e^{-{\rm i}k^+x^- -{\rm i}
\frac{2 \pi n^{\perp}}{L_\perp} x_\perp}  \right.    \nonumber \\
 & &  \left.
 +   a^\dagger_{ji}(k^+,n^{\perp})e^{{\rm i}k^+x^- +
{\rm i}\frac{2 \pi n^{\perp}}{L_\perp}  x_\perp}\right]\,,
\end{eqnarray}
\begin{eqnarray}
\psi_{ij}(0,x^-,x_\perp) &=&
   \frac{1}{2\sqrt{\pi L_\perp}}\sum_{n^{\perp}=-\infty}^{\infty}\int_0^\infty
       dk^+  
\left[b_{ij}(k^+,n^{\perp})e^{-{\rm i}k^+x^- -
{\rm i}\frac{2 \pi n^{\perp}}{L_\perp} x_\perp}  \right.   \nonumber \\
& & \left.   + b^\dagger_{ji}(k^+,n^\perp)e^{{\rm i}k^+x^- +{\rm i}
\frac{2 \pi n^{\perp}}{L_\perp} x_\perp}\right]\,,
\end{eqnarray}
where in (2+1) dimensions $n^\perp$ is the only transverse momentum index.

The chiral components of the supercharge are
\begin{eqnarray}\label{eq:supercharge+}
Q_{\rm SYM}^+&=&2^{1/4}\int dx^- dx_\perp
   \mbox{tr}\left[\phi\partial_-\psi-\psi\partial_-\phi\right]\,, \\
\label{eq:supercharge-}
Q_{\rm SYM}^-&=&2^{3/4}\int dx^- dx_\perp
       \mbox{tr}\left[\partial_\perp\phi\psi+
          g\left({\rm i}[\phi,\partial_-\phi]
             +2\psi\psi\right)\frac{1}{\partial_-}\psi\right]\,.
\end{eqnarray}
They satisfy the supersymmetry algebra
\begin{equation}
\{Q^+,Q^+\}=2\sqrt{2}P^+\,, \;\;
\{Q^-,Q^-\}=2\sqrt{2}P^-\,, \;\;
\{Q^+,Q^-\}=-4P_\perp\,. 
\end{equation}

This theory has the additional symmetries of transverse parity, $P$: 
$a_{ij}(k,n^\perp)\rightarrow -a_{ij}(k,-n^\perp)$,
$b_{ij}(k,n^\perp)\rightarrow b_{ij}(k,-n^\perp)$
and Kutasov's transposition\cite{Kutasov} $S$:
$a_{ij}(k,n^\perp)\rightarrow -a_{ji}(k,n^\perp)$,
$b_{ij}(k,n^\perp)\rightarrow -b_{ji}(k,n^\perp)$.
These allow the matrix representation to be block
diagonalized by an appropriate choice of basis.  Eigenstates
are labeled by the quantum numbers $\pm 1$ associated with $P$ and $S$.

\subsection{Spectrum and wave functions}

The main results for the spectrum of the SYM$_{2+1}$ theory
are given in Figs.~1, 11, and 12 of Ref.~\citebk{SYM2+1b}.
They show that the masses squared can be classified according to
three main forms of behavior: 
$1/L_\perp^2$, $g^2N_c\Lambda_\perp$, and $\Lambda_\perp^2$.
In particular, the spectrum as a function of $g$ separates
into two bands, one of approximately constant $M^2L_\perp^2$ and the
other growing rapidly.

For states in the lower band, the average number of constituents
increases rapidly with $g$.  At $g\simeq 1.5\sqrt{4\pi^3/N_cL_\perp}$
the DLCQ limit of $K$ constituents is saturated.  Thus
in SYM theory the low-mass states are dominated by Fock states with
many constituents, in close correspondence with string theory.
However, as a practical matter, the saturation means that the 
SDLCQ approximation breaks down, and numerical studies in this
band must be limited to smaller couplings.

Within the coupling limitation, extrapolations to infinite
resolution are easily done for low-mass states.  One first 
considers $M^2$ as a function of $1/T$ for a sequence of 
fixed $K$ values.  The extrapolations to $T=\infty$ then
yield $M^2$ as a function of $1/K$ alone, to extrapolate
to $K=\infty$.  The different representatives of continuum 
eigenstates are disentangled by studying their properties, such as
average constituent content and momentum.  Of course,
the different $P$ and $S$ symmetry sectors are explicitly
separated at the start.  Typical extrapolations are 
illustrated in Ref.~\citebk{SYM2+1b} with plots in
Figs.~4-8 and results in Tables~II and III.

For the spectrum as a whole there is a curious behavior
with respect to the average number of fermion constituents
$\langle n_F\rangle$.  Calculations for transverse resolution
$T=1$ and longitudinal resolutions $K=5$ and 6, and
for many different coupling strengths, show a gap between
$\langle n_F\rangle=4$ and 6, where no state is found.

Wave functions are also obtained in the diagonalization
process.  In the analysis of the spectrum they were
used to compute various average quantities that helped
identify states computed at different resolutions.  More
of the form of the wave function is revealed in the
structure function
\begin{eqnarray}
g_A(n,n^\perp)&=&\sum_{q=2}^K\sum_{n_1,\ldots,n_q=1}^{K-q}
\sum_{n^\perp_1,\ldots,n^\perp_q=-T}^{T} 
\delta\left(\sum_{i=1}^q n_i-K\right)
\delta\left(\sum_{j=1}^q n^{\perp}_j\right)  \nonumber \\
&&\qquad\times
\sum_{l=1}^q \delta^{n_l}_n\delta^{n^\perp_l}_{n^\perp}\delta^A_{A_l}
|\psi(n_1,n^\perp_1;\ldots; n_q,n^\perp_q)|^2\,,
\end{eqnarray}
where $A$ and $A_l$ represent the statistics (bosonic or fermionic) of the
probed type and the $l$-th constituent, respectively, and $\psi$ is a
Fock-state wave function.  In the lower band the shapes are
typically simple and are found to confirm the
identification of states at different resolutions.  In the
upper band, there are complicated shapes with multiple
bumps in transverse momentum, such as in Figs.~13 and 14 of 
Ref.~\citebk{SYM2+1b}.

\subsection{Stress-energy correlator}

Consider the following correlator of the stress-energy
component $T^{++}$: 
\begin{equation}
F(x^+,x^-,x^\perp) \equiv 
\langle 0| T^{++}(x^+,x^-,x^\perp) T^{++}(0,0,0)|0 \rangle\,,
\end{equation}
at strong coupling,\cite{T2_2+1} as an example of what one might compare 
with a supergravity 
approximation to string theory for small curvature.\cite{Maldacena}
In the discrete approximation, $F$ can be written as
\begin{eqnarray}
F(x^+,x^-,0)&=&\sum_{n,m,s,t} \,\,\left(\frac{\pi}{2L_\|^2 L_\perp}\right)^2 \\
&& \times \langle 0|\frac{L_\|}{\pi}T(n,m) e^{-iP^-_{op}x^+-iP^+x^-}
            \frac{L_\|}{\pi} T(s,t)| 0 \rangle\,,   \nonumber
\end{eqnarray}
where
\begin{eqnarray}
\frac{L_\|}{\pi}T^{++}(n,m) | 0 \rangle &=&
  {\sqrt{n m} \over 2}
    {\rm tr} \left[ a^\dagger_{ij}(n,n_\perp) a^\dagger_{ji} (m,m_\perp)
   \right]| 0\rangle  \\
&& + {(n-m) \over 4}
{\rm tr}\left[b^\dagger_{ij}(n,n_\perp) b^\dagger_{ji}(m,m_\perp)
\right] | 0 \rangle\,. \nonumber
\end{eqnarray}
Insertion of a complete set of bound states $|\alpha\rangle $
with light-cone energies $P_\alpha^-=(M_\alpha^2+P_\perp^2)/P^+$
at resolution K (and therefore $P^+=\pi K/L_\|$) and with total 
transverse momentum $P_\perp=2T\pi/L_\perp$ yields
\begin{eqnarray}
\frac{1}{\sqrt{-i} }\left(\frac{x^-}{x^+} \right)^2 F(x^+,x^-,0)
&=&\sum_\alpha 
\frac{1}{2 (2\pi)^{5/2}}\frac{M_\alpha^{9/2}}{\sqrt{r}}K_{9/2}(M_\alpha r)
\frac{|\langle u|\alpha\rangle|^2}{L_\perp K^3 |N_u|^2}\,, \nonumber \\
&& \mbox{ }
\label{eq:master}
\end{eqnarray}
with $r^2=x^+ x^-$, $x_\perp=0$, and
\begin{equation}
|u\rangle = N_u \frac{L_\|}{\pi}
  \sum_{n,m}\delta_{n+m,K}\delta_{n_\perp+m_\perp,N_\perp}T(n,m) |0 \rangle\,.
\end{equation}
Here $N_u$ is a normalization factor such that $\langle u|u\rangle=1$.
The sum over the full set of eigenvalues can be avoided by a
Lanczos-based technique.\cite{T2_1+1}

For free particles, $(x^+/x^-)^2F$ has a $1/r^6$ behavior.\cite{T2_2+1}
In the interacting case, this behavior should be recovered for small $r$,
where the bound states behave as free particles.  Because this
behavior depends on having a mass spectrum that extends to infinity,
the finite resolution of the numerical calculation yields only
$1/r^5$; however, the $1/r^6$ behavior is recovered in a careful limiting
process.

The behavior for large $r$ is determined by the massless states.
Because this theory has zero central charge, there are exactly
massless Bogomol'nyi--Prasad--Sommerfield (BPS) states at any
coupling.  However, their wave functions remain sensitive to
the coupling, and, at a particular (resolution dependent) value of $g$,
the correlator is exactly zero in the large-$r$ limit.  The
associated `critical' value of $g$ increases in proportion to
the square root of the transverse resolution $T$.\cite{T2_2+1}

\section{SYM-CS theory} \label{sec:SYM-CS}

The following CS term can be added to the Lagrangian:
\begin{equation}
L_{\rm CS}=\frac{\kappa}{2}\epsilon^{\mu\nu\lambda}
  \left(A_{\mu}\partial_{\nu}A_{\lambda}+\frac{2i}{3}gA_\mu A_\nu A_\lambda \right)
+\kappa\bar{\Psi}\Psi\,.
\end{equation}
This induces an additional term in the supercharge
\begin{equation}
\kappa Q_{\rm CS}^-\equiv-2^{3/4}\kappa \int dx^-                   
                \partial_-\phi\frac{1}{\partial_-}\psi\,,
\end{equation}
which generates in $P^-$ terms proportional to $\kappa^2$  
that act like a constituent
mass squared.  The presence of an effective mass reduces the 
tendency for low-mass states to be composed of high-multiplicity
Fock states.  This creates a theory in which the eigenstates are
more likely to be QCD-like, {\em i.e.}\ valence dominated, and
improves the applicability of the SDLCQ approximation to a
greater range of couplings.\cite{CSSYM1+1,BPS1+1,BPS2+1}

The dominance of the valence state is most prominent in the 
dimensionally reduced theory.\cite{CSSYM1+1}  This
(1+1)-dimensional theory is obtained by requiring the
fields to be constant in the transverse direction
and replacing $\partial_\perp$ by zero in the
full supercharge $Q_{\rm SYM}^- +\kappa Q_{\rm CS}^-$.  The
SYM contribution is then proportional to $g$.  Figure~3 of
Ref.~\citebk{CSSYM1+1} illustrates the dramatic reduction
in the average number of constituents as the ratio $\kappa/g$
is increased.  Also, there are states for which the mass
is nearly independent of $g$ at fixed $\kappa$.  These are
identified as approximate BPS states and are the reflection
of the massless BPS states in the underlying SYM theory.\cite{BPS1+1}
This behavior can be seen in Fig.~1 of Ref.~\citebk{BPS1+1}.

Similar anomalously light states appear in the full (2+1)-dimensional
theory.\cite{BPS2+1}  The bulk of the spectrum is driven to large
$M^2$ values as $g$ is increased, but one or more states remain
at low values.  The presence of the transverse degree of freedom
makes this more difficult to disentangle numerically, because
the matrices are larger and because one must consider the
transverse resolution limit.  Also, the eigenstates are less
valence-dominated at stronger YM coupling, which makes the
SDLCQ approximation less useful.  However, structure functions
have been extracted at intermediate coupling to show that
the approximate BPS states have a distinctly flat dependence
in longitudinal momentum.  Figure 3b of Ref.~\citebk{BPS2+1}
gives an example of this behavior.

\section{Summary} \label{sec:summary}

This work shows that one can compute spectra, wave functions,
and matrix elements nonperturbatively in supersymmetric theories.
The introduction of a Chern--Simons term brings an effective
constituent mass which has the effect of reducing the tendency
of SYM theory to form stringy, low-mass states with many constituents.
Instead, the lowest-mass states tend to be dominated by their
valence Fock state.
The massless BPS states of SYM theory survive in SYM-CS theory as states
with masses nearly independent of the YM coupling.

A number of interesting issues remain to be explored.
They include theories in 3+1 dimensions,
matter in the fundamental representation,\cite{FundMatter} and supersymmetry
breaking.  All of these are important for making 
contact with QCD.

\section*{Acknowledgments}
This talk was based on work done in collaboration with
S.S. Pinsky and U. Trittmann and supported in part 
by the U.S. Department of Energy and by grants 
of computing time from the Minnesota Supercomputing Institute.

\end{document}